\documentclass{Interspeech}

\interspeechcameraready 

\title{MVP: Multi-source Voice Pathology detection}

\author[affiliation={1}]{Alkis}{Koudounas*}
\author[affiliation={2}]{Moreno}{La Quatra*}
\author[affiliation={1}]{Gabriele}{Ciravegna}
\author[affiliation={3,4}]{Marco}{Fantini}
\author[affiliation={4}]{Erika}{Crosetti}
\author[affiliation={4,5}]{Giovanni}{Succo}
\author[affiliation={1}]{Tania}{Cerquitelli}
\author[affiliation={6}]{Sabato Marco}{Siniscalchi}
\author[affiliation={1}]{Elena}{Baralis}

\affiliation{}{Politecnico di Torino}{Italy}
\affiliation{}{Kore University of Enna}{Italy}
\affiliation{}{San Feliciano Hospital}{Italy\\}
\affiliation{}{Ospedale San Giovanni Bosco}{Italy}
\affiliation{}{Università degli Studi di Torino}{Italy\\}
\affiliation{}{Università degli Studi di Palermo}{Italy}

\email{alkis.koudounas@polito.it moreno.laquatra@unikore.it}

\usepackage{comment}
\usepackage{multirow}
\usepackage[normalem]{ulem}
\usepackage{pifont}
\useunder{\uline}{\ul}{}
\usepackage{cite}

\newcommand{\equalcontrib}{$^{*}$}
\newcommand\blfootnote[1]{%
  \begingroup
  \renewcommand\thefootnote{}\footnote{#1}%
  \addtocounter{footnote}{-1}%
  \endgroup
  }
  
\begin{document}

\maketitle

\begin{abstract}
     Voice disorders significantly impact patient quality of life, yet non-invasive automated diagnosis remains under-explored due to both the scarcity of pathological voice data, and the variability in recording sources. 
     This work introduces MVP (Multi-source Voice Pathology detection), a novel approach that leverages transformers operating directly on raw voice signals. 
     We explore three fusion strategies to combine sentence reading and sustained vowel recordings: waveform concatenation, intermediate feature fusion, and decision-level combination.
     Empirical validation across the German, Portuguese, and Italian languages shows that intermediate feature fusion using transformers best captures the complementary characteristics of both recording types.
     Our approach achieves up to +13\% AUC improvement over single-source methods.
\end{abstract}

\section{Introduction}
\blfootnote{\equalcontrib{} Both authors contributed equally to this work.}
Voice disorders affect approximately 30\% of the general population during their lifetime \cite{roy2004prevalence, roy2005voice, fantini2024rapidly}.
These conditions impact daily communication, work performance, and overall quality of life.
Disorders range from functional issues such as muscle tension dysphonia to organic pathologies such as vocal cord nodules \cite{akbulut2019basics}.
Early detection is crucial, as untreated disorders often progress to chronic conditions and cause psychological distress \cite{cohen2015delayed}.
Current clinical diagnosis relies on specialized equipment and clinicians. This approach is costly, invasive, and limited by specialist availability. There is thus a clear need for accessible, non-invasive screening methods for early detection.

Automated voice pathology detection offers a promising solution through machine learning analysis of voice recordings \cite{koudounas24_interspeech}.
Previous work has explored various approaches, including multilayer perceptrons trained on hand-crafted features such as Mel-frequency cepstral coefficients (MFCCs)~\cite{salhi2008voice, arias2018byovoz}, as well as convolutional neural networks (CNNs) applied to spectrograms~\cite{peng2023voice, xie2023voice}. Hybrid models combining CNNs and recurrent neural networks (RNNs) have also been investigated to better capture temporal dependencies in voice signals \cite{lilhore2023hybrid}. More recently, researchers have explored architectures that operate directly on raw audio, such as 1D-CNNs \cite{islam2022voice} and transformers \cite{ribas2023automatic}, with the latter showing promising results in both voice pathology and related speech disorders like dysarthria \cite{shahamiri2023dysarthric, almadhor2023e2e, ilias2023detecting, quatra_2024_exploiting, BDHPD}.

Current methods analyze either sustained vowels or continuous speech in isolation \cite{ribas2023automatic, almadhor2023e2e, quatra_2024_exploiting}.
Sustained vowels offer stable conditions for voice analysis but miss the dynamics of natural speech. Conversely, continuous speech captures everyday voice use and adds complexity through linguistic content and prosody.
Voice pathologies show different patterns across these speaking tasks, limiting the effectiveness of single-source analysis.
For instance, vocal nodules may affect sustained phonation more noticeably, while muscle tension disorders might be more evident during continuous speech~\cite{lee2019vocal, schlotthauer2010pattern}.

To address this limitation, we propose MVP (\textbf{M}ulti-source \textbf{V}oice \textbf{P}athology detection), a framework that leverages both sustained vowels and sentence readings through specialized transformer models \cite{hsu2021hubert, ARCH, voc2vec}, and allows a comprehensive assessment of vocal health. We specifically employ models pre-trained on LibriSpeech~\cite{panayotov2015librispeech} to analyze sentences and models pre-trained on AudioSet~\cite{gemmeke2017audio} to process sustained vowels.
Our method builds on recent advances in transformer-based voice pathology detection \cite{koudounas24_interspeech}, but while previous work exploited ensemble model separately trained on different sources, our approach explicitly explores multi-source integration at multiple levels.
We investigate three types of fusion strategy: waveform concatenation, intermediate representation fusion, and decision-level combination.
Each strategy addresses different aspects of multi-source integration, from raw signal combination to high-level feature fusion.
By employing specialized transformer architectures for each recording type, MVP captures their unique acoustic characteristics while maintaining crucial temporal relationships for pathology detection.
The contributions of this work are as follows.
\begin{itemize}
    \item \textit{Multi-source framework.} MVP combines sustained vowels and sentence readings through specialized architectures.
    \item \textit{Different fusion strategies}. We compare different strategies and show the advantages of intermediate representation fusion in this context.
    \item \textit{Extensive evaluation.} Experiments on three datasets demonstrate robust performance across different languages and recording conditions, with up to 13\% AUC improvement over state-of-the-art single-source methods.
\end{itemize}

\begin{figure}[ht!]
    \centering
    \includegraphics[width=\columnwidth]{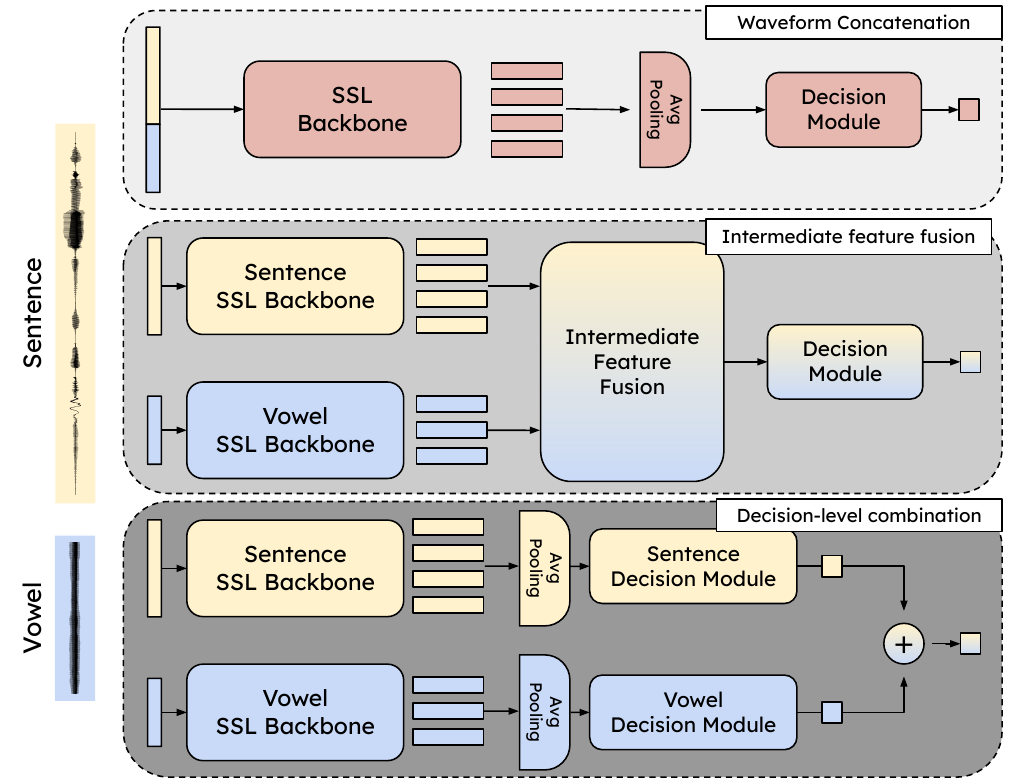}
    \caption{\textbf{Proposed MVP framework}:
    waveform concatenation (top panel), intermediate feature fusion (mid panel), and decision-level combination (bottom panel).}
    \vspace{-5mm}
    \label{fig:mvp_framework}
\end{figure}

\section{MVP Framework}
Our MVP framework addresses voice pathology detection by combining multiple recording sources.  
It adopts a three-stage architecture: (i) source-specific backbone models for feature extraction, (ii) fusion strategies to combine information from different sources, and (iii) a decision-making module for pathology detection.  
Fig.~\ref{fig:mvp_framework} provides an overview of the method.  

\subsection{Backbone Models}
We denote a sustained vowel recording as $X_{\text{SV}}$ and a sentence recording as $X_{\text{S}}$. 
For waveform concatenation, a single backbone processes the combined input.
For intermediate fusion and decision-level combination, each input is processed by specialized backbones.   
For sentence recordings, we use HuBERT pre-trained on LibriSpeech \cite{panayotov2015librispeech}, which captures linguistic content and prosodic variations.  
For sustained vowel recordings, we use HuBERT pre-trained on AudioSet \cite{gemmeke2017audio}, focusing on vocal quality and non-semantic features.  
The frame-level representations extracted by each backbone are:  
\begin{equation}
    H_{\text{SV}} = \text{HuBERT}_{\text{AS}}(X_{\text{SV}}), \quad H_{\text{S}} = \text{HuBERT}_{\text{LS}}(X_{\text{S}})
\end{equation}
where $H_{\text{SV}} \in \mathbb{R}^{T_{\text{SV}} \times d}$ and $H_{\text{S}} \in \mathbb{R}^{T_{\text{S}} \times d}$ contain temporal information from each recording type, $T_{\text{SV}}$, $T_{\text{S}}$ represents the source lengths
and $d$ is the latent dimensionality of the representations.  

\subsection{Fusion Strategies}
\label{subsec:fusion_strategies}
We explore three strategies for combining information from sustained vowel and sentence recordings, each operating at a different level of abstraction.

\begin{itemize}
    \item \textit{Waveform Concatenation (WC).}
    Raw audio signals are concatenated and processed by a single backbone, preserving all information but potentially introducing artifacts due to the different characteristics of each source (Fig. \ref{fig:mvp_framework} (top)).
    
    \item \textit{Intermediate Feature Fusion (IFF).}
    Features from specialized backbones are combined through fusion methods (Section \ref{subsec:fusion_methods}).
    This preserves source-specific characteristics while enabling cross-source learning (Fig. \ref{fig:mvp_framework} (middle)).
    
    \item \textit{Decision-Level Combination (DLC).}
    This approach mimics the pipeline proposed in~\cite{koudounas24_interspeech}, where an ensemble dynamically selects between source-specific predictions by averaging individual backbones probabilities (Fig. \ref{fig:mvp_framework} (bottom)). 
\end{itemize}

\subsubsection{Intermediate Feature Fusion}
\label{subsec:fusion_methods}
Given two feature sequences $H_{\text{SV}} \in \mathbb{R}^{T_{\text{SV}} \times d}$ and $H_{\text{S}} \in \mathbb{R}^{T_{\text{S}} \times d}$ from sustained vowels and sentence recordings, respectively, we explore five methods to fuse them into a single vector $Z_{\text{fused}} \in \mathbb{R}^d$.
Each method represents a different approach to capturing cross-source interactions~\cite{LaQuatra2024}.

\vspace{1mm}
\noindent \textbf{Simple Concatenation (Baseline).}
We first apply average pooling to each sequence independently to obtain global representations $Z_{\text{SV}}^{\text{Avg}}$ and $Z_{\text{S}}^{\text{Avg}}$ for each source.
The fused representation is their concatenation:
\begin{equation}
    Z_{\text{Concat}} = [Z_{\text{SV}}^{\text{Avg}}; Z_{\text{S}}^{\text{Avg}}] \in \mathbb{R}^{2d}
\end{equation}
This approach preserves the global characteristics of each source while providing a strong baseline for comparison.

\vspace{1mm}
\noindent \textbf{Attention Pooling (AP).}
We concatenate sequences on the time dimension and apply a learned attention mechanism to capture the relative importance of different time steps.
The attention scores $\alpha_t$ are computed using a learnable vector $w \in \mathbb{R}^d$:
\begin{equation}
    \alpha_t = \frac{\exp(w^\top H_t)}{\sum_{t'} \exp(w^\top H_{t'})}, \quad Z_{\text{AP}} = \sum_t \alpha_t H_t \in \mathbb{R}^d
\end{equation}
where $H_t$ represents features at time step $t$ from the concatenated sequence $[H_{\text{SV}}; H_{\text{S}}]$.
This method emphasizes the most informative temporal features across both sources.

\vspace{1mm}
\noindent \textbf{Gating Mechanism.}
This method introduces adaptive weighting between sources through a learned gating mechanism \cite{arevalo2017gated}.
We first obtain source-specific vectors using AP, then compute an adaptive gating vector:
\begin{equation}
    G = \sigma(W [Z_{\text{SV}}^{\text{AP}}; Z_{\text{S}}^{\text{AP}}]), \quad Z_{\text{Gating}} = G \odot [Z_{\text{SV}}^{\text{AP}}; Z_{\text{S}}^{\text{AP}}]
\end{equation}
where $W$ is a learnable matrix and $\sigma$ is the sigmoid function.
The gating allows the model to dynamically adjust the contribution of each source.

\vspace{1mm}
\noindent \textbf{Feature-wise Linear Modulation (FiLM).}
FiLM enables bidirectional cross-source interactions by allowing each source to modulate the other \cite{perez2018film}:
\begin{equation}
    Z_{\text{S}}^{\text{FiLM}} = Z_{\text{S}}^{\text{AP}} \odot (1 + W_\gamma^s Z_{\text{SV}}^{\text{AP}}) + W_\beta^s Z_{\text{SV}}^{\text{AP}}
\end{equation}
\begin{equation}
    Z_{\text{SV}}^{\text{FiLM}} = Z_{\text{SV}}^{\text{AP}} \odot (1 + W_\gamma^v Z_{\text{S}}^{\text{AP}}) + W_\beta^v Z_{\text{S}}^{\text{AP}}
\end{equation}
where $W_\gamma^s, W_\gamma^v, W_\beta^s, W_\beta^v$ are learnable parameters for scale and shift operations.
The final representation combines both modulated features:
\begin{equation}
    Z_{\text{FiLM}} = [Z_{\text{S}}^{\text{FiLM}}; Z_{\text{SV}}^{\text{FiLM}}]
\end{equation}
This bidirectional modulation allows each source to influence the other through learned transformations.

\vspace{1mm}
\noindent \textbf{Transformer Encoder (TE).}
We first concatenate the sequences along the time axis to form a combined sequence:
\begin{equation}
    H_{\text{combined}} = [H_{\text{SV}}; H_{\text{S}}] \in \mathbb{R}^{(T_{\text{SV}} + T_{\text{S}}) \times d}
\end{equation}
This sequence is processed through $L$ transformer encoder layers, where self-attention mechanisms enable each time step from one source to interact with all the time steps from both sources.
This fine-grained interaction preserves and leverages the temporal structure of both sources.
The final representation is obtained through attention pooling:
\begin{equation}
    Z_{\text{TE}} = \text{AP}(\text{TE}_L(H_{\text{combined}})) \in \mathbb{R}^d
\end{equation}

\subsection{Decision-Making Module}
The output from any of the fusion strategies results in a final fused representation $Z$.  
This representation is passed to the decision-making module, which consists of a fully connected layer followed by a sigmoid activation function:
\begin{equation}
    \hat{y} = \text{Sigmoid}(\text{FC}(Z))
\end{equation}

\noindent
The output $\hat{y} \in [0, 1]$ represents the probability of the positive class (presence of voice pathology).

\begin{table*}[h!]
\centering
\caption{\textbf{Mean\scriptsize±std \normalsize performance of different fusion approaches across three datasets}. Best results in \textbf{bold}, second-best \underline{underlined}. Models use either LibriSpeech (\texttt{LS}) or AudioSet (\texttt{AS}) pre-training.
\texttt{$\rightarrow$Sent} indicates fine-tuning on read sentences, \texttt{$\rightarrow$Vowel} on sustained vowels, \texttt{$\rightarrow$Mix} on both sources. \ding{100} indicates frozen backbones. IFF fusion is implemented with TE.}
\vspace{-3mm}
\renewcommand{\arraystretch}{0.91}
\label{tab:acc}
\resizebox{\textwidth}{!}{%
\begin{tabular}{l@{\hspace{0.5em}}c@{\hspace{0.5em}}|ccc|ccc|ccc}
\toprule
\multirow{2.5}{*}{\textbf{Model}} & 
\multirow{2.5}{*}{\textbf{\# Params}} & 
\multicolumn{3}{c}{\textbf{SVD}} & 
\multicolumn{3}{c}{\textbf{AVFAD}} & 
\multicolumn{3}{c}{\textbf{IPV}} \\
\cmidrule{3-11}
& & \textbf{Acc.} & \textbf{F1} & \textbf{AUC} & \textbf{Acc.} & \textbf{F1} & \textbf{AUC} & \textbf{Acc.} & \textbf{F1} & \textbf{AUC} \\
\midrule
\multicolumn{11}{c}{\textit{Single-Source Baselines}} \\ \midrule

\texttt{LS$\rightarrow$Sent}
    & 94.64M                                      
    & .873\scriptsize±.058          
    & .849\scriptsize±.062          
    & .850\scriptsize±.048          
    & .872\scriptsize±.015          
    & .871\scriptsize±.014          
    & .877\scriptsize±.015          
    & .875\scriptsize±.024          
    & .870\scriptsize±.026          
    & .847\scriptsize±.026 \\
    
\texttt{LS$\rightarrow$Vowel} 
    & 94.64M                                      
    & .747\scriptsize±.075          
    & .724\scriptsize±.074          
    & .732\scriptsize±.084          
    & .714\scriptsize±.051          
    & .705\scriptsize±.061          
    & .714\scriptsize±.061          
    & .622\scriptsize±.064          
    & .617\scriptsize±.062          
    & .620\scriptsize±.062  \\
    
\texttt{AS$\rightarrow$Sent}
    & 94.64M                                      
    & .817\scriptsize±.060          
    & .801\scriptsize±.061          
    & .810\scriptsize±.052          
    & .852\scriptsize±.045          
    & .850\scriptsize±.047          
    & .855\scriptsize±.050          
    & .828\scriptsize±.039          
    & .823\scriptsize±.038          
    & .815\scriptsize±.044 \\

\texttt{AS$\rightarrow$Vowel} 
    & 94.64M                                      
    & .779\scriptsize±.075          
    & .747\scriptsize±.068          
    & .760\scriptsize±.079          
    & .798\scriptsize±.043          
    & .758\scriptsize±.056          
    & .756\scriptsize±.058          
    & .676\scriptsize±.059          
    & .649\scriptsize±.060          
    & .665\scriptsize±.055 \\
 
\texttt{LS$\rightarrow$Mix} 
    & 94.64M                      
    & .780\scriptsize±.028         
    & .791\scriptsize±.031          
    & .765\scriptsize±.025          
    & .827\scriptsize±.017    
    & .826\scriptsize±.019
    & .827\scriptsize±.019         
    & .840\scriptsize±.018          
    & .831\scriptsize±.015          
    & .831\scriptsize±.016 \\ \midrule
    
\multicolumn{11}{c}{\textit{Waveform Concatenation (WC)}} \\ \midrule

\texttt{LS}
    & 94.64M                                      
    & .896\scriptsize±.054          
    & .882\scriptsize±.053          
    & .891\scriptsize±.056          
    & \underline{.907\scriptsize±.054}          
    & \underline{.906\scriptsize±.055}          
    & \underline{.908\scriptsize±.052}          
    & .888\scriptsize±.066          
    & \underline{.881\scriptsize±.067}         
    & \underline{.886\scriptsize±.060} \\

\texttt{AS} 
    & 94.64M                                      
    & .875\scriptsize±.063          
    & .869\scriptsize±.061          
    & .873\scriptsize±.068          
    & .889\scriptsize±.042          
    & .884\scriptsize±.045          
    & .885\scriptsize±.049          
    & .836\scriptsize±.031          
    & .832\scriptsize±.032          
    & .831\scriptsize±.037 \\
    \midrule

\multicolumn{11}{c}{\textit{Intermediate Feature Fusion (IFF)}} \\ \midrule

\texttt{LS+AS} \ding{100}
    & 17.98M                                      
    & .826\scriptsize±.036          
    & .809\scriptsize±.035          
    & .832\scriptsize±.023          
    & .833\scriptsize±.032          
    & .831\scriptsize±.032          
    & .834\scriptsize±.033          
    & .813\scriptsize±.048          
    & .806\scriptsize±.045          
    & .809\scriptsize±.048 \\

\texttt{LS+AS}
    & 206.73M                            
    & \textbf{.958\scriptsize±.063} 
    & \textbf{.953\scriptsize±.067} 
    & \textbf{.958\scriptsize±.062} 
    & \textbf{.962\scriptsize±.040} 
    & \textbf{.962\scriptsize±.039} 
    & \textbf{.963\scriptsize±.038} 
    & \textbf{.939\scriptsize±.044} 
    & \textbf{.931\scriptsize±.054} 
    & \textbf{.936\scriptsize±.053} \\
    \midrule

\multicolumn{11}{c}{\textit{Decision-Level Combination (DLC)}} \\ \midrule

\texttt{LS+LS} 
    & 189.28M                                     
    & .885\scriptsize±.062          
    & .863\scriptsize±.070          
    & .860\scriptsize±.070          
    & .881\scriptsize±.095          
    & .874\scriptsize±.111          
    & .879\scriptsize±.099          
    & .882\scriptsize±.083          
    & .873\scriptsize±.084          
    & .862\scriptsize±.064 \\

\texttt{AS+AS}
    & 189.28M                                     
    & .864\scriptsize±.072          
    & .855\scriptsize±.076          
    & .857\scriptsize±.075          
    & .872\scriptsize±.034          
    & .863\scriptsize±.034          
    & .866\scriptsize±.035          
    & .837\scriptsize±.061          
    & .826\scriptsize±.095          
    & .827\scriptsize±.102 \\

\texttt{LS+AS} 
    & 189.28M                                     
    & {\ul .898\scriptsize±.051}    
    & {\ul .884\scriptsize±.058}    
    & {\ul .896\scriptsize±.058}    
    & .888\scriptsize±.092    
    & .887\scriptsize±.108
    & .889\scriptsize±.096
    & {\ul .896\scriptsize±.072}    
    & .877\scriptsize±.074
    & .882\scriptsize±.062 \\   
    \bottomrule

\end{tabular}
}
\vspace{-3mm}
\end{table*}

\section{Experimental Setup}

To ensure reproducibility and fair comparison, in the following, we detail our experimental setup across datasets, training, data processing, and evaluation procedures\footnote{\scriptsize{\url{https://github.com/koudounasalkis/MVP}}}.

\subsection{Datasets}
We evaluate our approach on three datasets that contain both sentence readings and sustained vowel emissions recorded under controlled conditions. 

\vspace{1mm}
\noindent \textbf{SVD.} The Saarbruecken Voice Database contains German voice recordings from 2,043 subjects (687 healthy and 1,356 pathological).
Each recording session includes sustained vowels at different pitch levels and a sentence-reading task.
For consistency with other datasets, we only use normal pitch vowels and sentence readings.

\vspace{1mm}
\noindent \textbf{AVFAD.} The Advanced Voice Function Assessment Database includes Portuguese recordings from 709 subjects (346 healthy and 363 pathological).
Each subject performs multiple tasks, including sustained vowels \texttt{/a/}, \texttt{/e/}, \texttt{/o/}, and reading six phonetically balanced sentences.
We randomly select one of the six sentences and one of the vowels for each subject.

\vspace{1mm}
\noindent \textbf{IPV.} The Italian Pathological Voice is a dataset including recordings from 513 subjects (173 healthy and 340 pathological). Voice samples were recorded under standardized conditions with a 30~cm mouth-to-microphone distance and ambient noise below 30 dB. Each subject recorded a sustained vowel \texttt{/a/} and five phonetically balanced sentences. We randomly select one sentence and the vowel for each speaker.

\subsection{Implementation Details}

\noindent
\textbf{Training Protocol and Data Processing.}
We perform 10-fold cross-validation across all experiments, ensuring speaker-independent splits. Each fold maintains the same healthy-to-pathological ratio as the original dataset.
All audio files are resampled to 16 kHz and normalized to zero mean and unit variance. For consistent processing, recordings are padded or truncated to a fixed length of 5.0 seconds. We use an AdamW optimizer with 5e-5 learning rate and 0.01 weight decay.
Training runs for 10 epochs with early stopping (patience=5) on validation loss.
We use binary cross-entropy loss and batch size 64.
Training was performed on a single NVIDIA A100 80GB GPU.
For the TE fusion strategy, we set $L=2$ transformer encoder layers, which provides an effective balance between model complexity and representational power.

\vspace{1mm}
\noindent \textbf{Data Augmentation.} 
We implement separate augmentation strategies for sentences and sustained vowels to preserve their distinct characteristics.
For sentence readings, we apply augmentation with 25\% probability, including noise addition (SNR between 0-30dB), speed perturbation (0.75x to 1.25x), pitch shifting (±4 semitones), and their combinations.
For sustained vowels, we apply augmentation with a lower probability of 10\% to preserve their core vocal characteristics.
This empirically-determined approach aims to balance data diversity with signal integrity, which is particularly important for sustained vowels where stable phonation is essential for pathology detection.

\vspace{1mm}
\noindent \textbf{Single-Source Baselines.}
We implement single-source baselines using HuBERT models pre-trained on either LibriSpeech (LS) or AudioSet (AS) as they both were proven effective in previous work~\cite{koudounas24_interspeech}.
Each baseline uses the same architecture as our multi-source models but processes only one recording type, enabling fair comparison of the multi-source advantage. An additional baseline involves a single HuBERT (LS) model trained on a mix of sentences and vowels. 

\vspace{1mm}
\noindent \textbf{Model Configuration and Evaluation.}
Our transformer backbone uses HuBERT in its base configuration with 94.64M parameters.
For single-source baselines, we evaluate both pre-trained models on each type of recording and the HuBERT (LS) model trained on the mix of sources.
In the IFF strategy, we experiment with both frozen backbones (\ding{100} -- 17.98M trainable parameters) and fine-tuned backbones (206.73M parameters). 
We apply the fusion strategy extracting representation from the 5th layer of the SSL backbones -- see Section \ref{subsec:ablations} for detailed ablation studies. 
For DLC, we train two LS- or AS- backbone models separately on sentences and vowels or a combined approach using both specialized backbones (LS+AS). In all cases, the number of trainable parameters is 189.28M.
We report accuracy, macro F1 score, and AUC-ROC averaged across folds with standard deviations to evaluate model performance.

\begin{table}[t]
    \caption{\textbf{A comparison of IFF fusion strategies,} AUC scores. Best results in \textbf{bold} second-best \underline{underlined}.}
    \vspace{-2mm}
    \small
    \centering
    \label{tab:ablation_fusion_mod}

    \begin{tabular}{l|ccc}
        \toprule
        \textbf{Method} & \textbf{SVD} & \textbf{AVFAD} & \textbf{IPV} \\
        \midrule
        \texttt{Concat}     & .918\scriptsize±.060 & .920\scriptsize±.044 & .915\scriptsize±.050 \\
        \texttt{AP}  & .948\scriptsize±.060 & .955\scriptsize±.040 & .929\scriptsize±.047 \\
        \texttt{TE}         & \textbf{.958\scriptsize±.062} & \textbf{.963\scriptsize±.038} & \textbf{.936\scriptsize±.053} \\
        \texttt{Gating}     & .947\scriptsize±.068 & .956\scriptsize±.045 & .926\scriptsize±.049 \\
        \texttt{FiLM}       & \underline{.951\scriptsize±.062} & \underline{.961\scriptsize±.041} & \underline{.934\scriptsize±.053} \\
        \bottomrule
    \end{tabular}
\vspace{-3mm}
\end{table}

\begin{table}[t]
    \caption{\textbf{Ablation study on backbone models,} AUC scores.
    \texttt{H}=HuBERT, \texttt{v2v}=voc2vec, \texttt{LS}=LibriSpeech, \texttt{AS}=AudioSet. Best results in \textbf{bold} second-best \underline{underlined}.}
    \vspace{-2mm}
    \small
    \label{tab:ablation_model}
    \resizebox{\columnwidth}{!}{%
    \begin{tabular}{ll|ccc}
        \toprule
        \textbf{Sentence} & \textbf{Vowel} & \textbf{SVD} & \textbf{AVFAD} & \textbf{IPV} \\
        \midrule
        \texttt{H-LS} & \texttt{H-LS}      & .942\scriptsize±.059 & .951\scriptsize±.042 & .917\scriptsize±.067 \\
        \texttt{H-AS} & \texttt{H-AS}      & .935\scriptsize±.072 & .939\scriptsize±.042 & .901\scriptsize±.048 \\
        \texttt{H-LS} & \texttt{H-AS}      & \textbf{.958\scriptsize±.062} & \textbf{.963\scriptsize±.038} & \textbf{.936\scriptsize±.053} \\
        \texttt{H-LS} & \texttt{v2v}    & \underline{.953\scriptsize±.056} & \underline{.958\scriptsize±.044} & \underline{.929\scriptsize±.044} \\
        \bottomrule
    \end{tabular}
    \vspace{-5mm}}
\end{table}

\begin{table}[t]
    \caption{\textbf{Ablation study on feature extraction layer depth}, AUC scores. Best results in \textbf{bold} second-best \underline{underlined}.}
    \vspace{-2mm}
    \small
    \centering
    \renewcommand{\arraystretch}{0.9}
    \label{tab:ablation_layer}
    \begin{tabular}{l|ccc}
        \toprule
        \textbf{Layer} & \textbf{SVD} & \textbf{AVFAD} & \textbf{IPV} \\
        \midrule
        \texttt{4th}       & .944\scriptsize±.054 & .945\scriptsize±.043 & .924\scriptsize±.048 \\
        \texttt{5th}       & \textbf{.958\scriptsize±.062} & \textbf{.963\scriptsize±.038} & \textbf{.936\scriptsize±.053} \\
        \texttt{6th}       & .950\scriptsize±.055 & .943\scriptsize±.046 & .923\scriptsize±.060 \\
        \texttt{7th}       & .948\scriptsize±.066 & .954\scriptsize±.042 & .929\scriptsize±.061 \\
        \texttt{Last}      & \underline{.954\scriptsize±.061} & \underline{.958\scriptsize±.037} & \underline{.932\scriptsize±.072} \\
        \texttt{Weighted}  & .953\scriptsize±.059 & .957\scriptsize±.042 & \underline{.932\scriptsize±.057} \\
        \bottomrule
    \end{tabular}
    \vspace{-5mm}
\end{table}

\section{Experimental Analysis}

Table \ref{tab:acc} demonstrates that our multi-source approach significantly outperforms single-source baselines across all datasets.
The IFF-TE method with fine-tuned backbones achieves the highest AUC scores: 95.8\% (SVD), 96.3\% (AVFAD), and 93.6\% (IPV).
This represents a 10-13\% improvement over the best single-source baseline, showing the clear advantage of the multi-source approach.
We separately investigate the impact of IFF fusion strategies in Section \ref{subsec:fusion_strategies_results}.

When focusing on the single-source baselines, the results confirm previous findings \cite{koudounas24_interspeech}: models consistently perform better on sentence readings compared to sustained vowels.
Sentence readings may contain richer diagnostic information, possibly because they capture both phonation quality and dynamic speech characteristics. 
However, training a single model on a mixture of sources proves ineffective,
This is likely because the model cannot adapt to the diversity of recording types, leading to weak overall performance.  
In contrast, the significant performance improvement observed with our multi-source approach highlights the complementary value of sustained vowels. 
IFF also outperforms other fusion strategies such as WC and DLC.  
Interestingly, WC performs better than DLC in two out of three datasets while requiring only a single model, actually halving the number of parameters.  
Even in simpler settings, concurrent access to both sources allows WC to learn cross-source patterns, supporting the value of joint analysis.
DLC exhibits high standard deviations, likely due to the limited amount of data, especially when performing 10-fold cross-validation.
An important finding also emerges from the resource-constrained version of IFF.
Although the best results are achieved by fine-tuning both backbones (206.73M parameters), the frozen variant (\ding{100}) still outperforms four of five single-source baselines with less than 20\% of their trainable parameters.
This highlights the benefits of multi-source analysis even in resource-constrained scenarios where fine-tuning large models may not be feasible.

\vspace{-1mm}
\subsection{Fusion Strategies}
\label{subsec:fusion_strategies_results}
\vspace{-1mm}

In Table \ref{tab:ablation_fusion_mod} we analyze the impact of different IFF strategies for cross-source learning.
The Transformer Encoder (TE) achieves the best overall performance, particularly on SVD and AVFAD, likely due to its ability to correlate temporal relationships between sources.
While FiLM and AP provide strong alternatives, the significant performance gap between learned fusion strategies and simple concatenation (up to 4.7\% on AVFAD) highlights the importance of modeling fine-grained interactions between sources.
Effective voice pathology detection requires careful modeling of how different vocal characteristics manifest across both sustained and dynamic speech patterns.
The results suggest these characteristics in isolation are insufficient; their correlations provide crucial diagnostic information.

\vspace{-1mm}
\subsection{Ablation Studies}
\label{subsec:ablations}
\vspace{-1mm}
To validate our choices, we conduct specific ablation studies examining: 
(i) the impact of model architecture and pre-training data sources, 
(ii) the optimal layer for feature extraction. 

\vspace{1mm}
\noindent
\textbf{Impact of model architectures and pre-training sources.}
Table \ref{tab:ablation_model} reveals the role of the model architecture and specialized pre-training for each source.
The combination of HuBERT models pre-trained on LibriSpeech (LS) for sentences and on AudioSet (AS) for sustained vowels consistently outperforms other configurations across all datasets.
This aligns with recent works~\cite{koudounas24_interspeech} showing that sentence analysis benefits from pre-training on speech data (LS), while sustained vowel analysis benefits from exposure to diverse acoustic events (AS).
To process sustained vowels, we also evaluate voc2vec~\cite{voc2vec} which mimics the wav2vec 2.0~\cite{wav2vec2} architecture and it is pre-trained on non-speech vocalizations.
It shows competitive performance without surpassing the HuBERT LS-AS combination.
Specialized pre-training can thus improve performance, but the acoustic event coverage in AudioSet provides more effective representations to capture pathological voice characteristics.  

\vspace{1mm}
\noindent
\textbf{Optimal Representations.}
The results in Table \ref{tab:ablation_layer} provide insights into the ideal point of feature extraction.
Although performance remains relatively stable across different layers, feature extraction at the 5th layer consistently gives optimal results.
Mid-level representations provide the best balance between preserving source-specific characteristics and enabling effective cross-source integration \cite{nguyen2024exploring}.
Learned weighted sum across all layers shows competitive but not superior performance, proving that it may need more data for optimal weighting. 

\section{Conclusions}
This paper presented MVP, a novel multi-source approach for voice pathology detection that effectively combines sustained vowels and sentence readings.
Our experimental results across three languages demonstrate that intermediate feature fusion with transformers consistently outperforms single-source methods, achieving up to 13\% AUC improvement.

\section{Acknowledgements}
This work is partially supported by the FAIR - Future Artificial Intelligence Research and received funding from the European Union Next-GenerationEU (PIANO NAZIONALE DI RIPRESA E RESILIENZA (PNRR) – MISSIONE 4 COMPONENTE 2, INVESTIMENTO 1.3 – D.D. 1555 11/10/2022, PE00000013), is partially supported by the "D.A.R.E. – Digital Lifelong Prevention" project (code: PNC0000002, CUP: B53C22006450001), co-funded by the Italian Complementary National Plan PNC-I.1 Research initiatives for innovative technologies and pathways in the health and welfare sector (D.D. 931 of 06/06/2022), and partially supported by the European Union – Next Generation EU under the National Recovery and Resilience Plan (PNRR) – M4 C2, Investment 1.1: Fondo per il Programma Nazionale di Ricerca e Progetti di Rilevante Interesse Nazionale (PRIN) - PRIN 2022 - "SHAPE-AD" (CUP: J53D23007240008).
This manuscript reflects only the authors' views and opinions, neither the European Union nor the European Commission can be considered responsible for them. 

\bibliographystyle{IEEEtran}
\bibliography{mybib}

\end{document}